\documentclass[preprint,showpacs,preprintnumbers,amsmath,amssymb]{revtex4}

\usepackage{graphicx}
\usepackage{dcolumn}
\usepackage{bm}

\newcommand{\vc}[1]{{\bm{#1}}}
\begin{document}

\title{Probability distributions of turbulent energy}  
\author{Mahdi Momeni}
\affiliation{Faculty of Physics, Tabriz University, Tabriz 51664, Iran}
\author{Wolf-Christian M\"uller}
\email{Wolf.Mueller@ipp.mpg.de}
\affiliation{Max-Planck-Institut f\"ur Plasmaphysik, 85748 Garching, Germany}
\date{\today}
\begin{abstract}
Probability density functions (PDFs) of scale-dependent energy
fluctuations, $P[\delta E(\ell)]$, are studied in high-resolution
direct numerical simulations of Navier-Stokes and incompressible
magnetohydrodynamic (MHD) turbulence. MHD flows with and without a
strong mean magnetic field are considered.  For all three systems it
is found that the PDFs of inertial range energy fluctuations exhibit
self-similarity and monoscaling in agreement with recent solar-wind measurements
[B. Hnat et al., Geophys. Res. Lett. 29(10), 86-1 (2002)].
Furthermore, the energy PDFs exhibit similarity over
all scales of the turbulent system showing no substantial qualitative
change of shape as the scale of the fluctuations varies. This is in
contrast to the well-known behavior of PDFs of turbulent
velocity fluctuations.  
In all three cases under consideration the
$P[\delta E(\ell)]$ resemble L\'evy-type gamma distributions $\sim
\Delta^{-1}\exp(-|\delta E|/\Delta)|\delta E|^{-\gamma}$  
The observed gamma distributions
exhibit a scale-dependent width $\Delta(\ell)$ and a system-dependent $\gamma$. 
The monoscaling property reflects the inertial-range scaling of the Els\"asser-field fluctuations due to lacking Galilei invariance of $\delta E$. The appearance of L\'evy distributions is made plausible by a simple model of energy transfer. 
\end{abstract}
\maketitle
%
Turbulence in electrically conducting magnetofluids is, apart from its
importance for laboratory plasmas (see, for example
\cite{ortolani_schnack:rfpbook}), a key ingredient in the dynamics of,
e.g., the earth's liquid core and the solar wind (see
e.g. \cite{biskamp:book3}).  
A simple description of such plasmas 
is the framework of incompressible magnetohydrodynamics (MHD), a fluid
approximation neglecting kinetic processes occuring on microscopic
scales. This approach is appropriate if the main interest is focused
on the nonlinear dynamics and the inherent statistical
properties of fluid turbulence.  
To this end, two-point increments of a turbulent field component, say $f$, in the
direction of a fixed unit vector $\mathbf{\hat{e}}$, $\delta
f(\ell)=f(\vc{r} + \mathbf{\hat{e}} \ell)-f(\vc{r})$ are analyzed, since they yield a
comprehensive and scale-dependent characterization of the statistical properties of
turbulent fluctuations via the associated probability density function (PDF)
\cite{monin_yaglom:book2}.

PDFs of
temporal fluctuations
\footnote{In the specific configuration time
scales can be linearly related to spatial scales (Taylor's hypothesis).}  
in the solar wind, e.g. of total (magnetic + kinetic)
energy density, as measured by the WIND spacecraft are self-similar over all
observed scales, exhibit monoscaling, and closely resemble gamma distributions. 
In contrast the PDFs of velocity and magnetic field are
found to display well-known multifractal characteristics, i.e. the
associated PDFs change from Gaussian at large scales to leptocurtic
(fat-tailed) at small scales
\cite{hnat_etal:monoscalsolar,hnat_chapman_rowlands:monoscalfokker,
hnat_chapman_rowlands:monoscalall}.  The solar wind plasma is
a complex and inhomogeneous mixture of mutually interacting
regions with different physical characteristics and dynamically
important kinetic processes \cite{tu_marsch:book,goldstein_etal:solwindrev}. Thus, it is not
clear if the abovementioned solar-wind observations are caused by turbulence or some other
physical phenomenon.  This paper reports an investigation of
turbulent PDFs based on high-resolution direct numerical simulations of
physically `simpler' homogeneous incompressible MHD and Navier-Stokes
turbulence to elucidate this question.  Monoscaling of the two-point
PDFs of energy is found in the inertial range of macroscopically
isotropic MHD turbulence, anisotropic MHD turbulence with an imposed
mean magnetic field as well as in turbulent Navier-Stokes flow.  The
respective PDFs resemble leptocurtic gamma laws on all spatial scales 
in agreement with the solar-wind measurements.
The monoscaling property is shown to be a consequence of
lacking Galilei invariance of the energy fluctuations in combination with turbulent inertial-range scaling.
The appearance of L\'evy-type gamma distributions apparently results from nonlinear 
turbulent tranfer
as suggested by similar findings in all three investigated systems and a simple reaction-rate model.
    
The dimensionless equations of incompressible MHD, formulated in
Els\"asser variables $\vc{z}^\pm=\vc{v}\pm\vc{b}$ with the fluid velocity $\vc{v}$ and the magnetic field $\vc{b}$ which
is given
in Alfv\'en-speed units \cite{elsaesser:mhd}, read
\begin{eqnarray}
 \nabla\cdot\vc{z}^\pm&=&0 \label{elsaesser1}\\ 
 \partial _{t} \vc{z}^{\pm}&=&-\vc{z}^{\mp}\cdot \nabla \vc{z}^{\pm}-\nabla P+\eta _{+} 
         \Delta \vc{z}^{\pm}+ \eta _{-} \Delta\vc{z}^{\mp}  \label{elsaesser2}
 \end{eqnarray}
with the total pressure $P=p+\frac {1}{2}b^{2}$. The dimensionless
kinematic viscosity $\mu$ and magnetic diffusivity $\eta$ appear in
$\eta_\pm=1/2(\mu\pm\eta)$. 

The data used in this work stems from
pseudospectral high-resolution direct numerical simulations
\cite{mueller_grappin:endyn} based on a set of equations equivalent
to Eqs. (\ref{elsaesser1}) and (\ref{elsaesser2}).  It describes
homogeneous fully-developed turbulent MHD and Navier-Stokes ($\vc{b}\equiv 0$) flows in a cubic box of
linear size $2\pi$ with periodic boundary conditions.
The initial conditions for the decaying simulation run consist of random fluctuations 
with total energy equal to unity. In the 
MHD cases total kinetic and magnetic energy are approximately equal. The initial spectral
energy distribution is peaked at small wavenumbers around $k=4$ and decreases like a Gaussian towards 
small scales. 
In the MHD setups magnetic and
cross helicity are small implying
$\vc{z}^{+}\simeq \vc{z}^{-}$. 
The driven turbulence simulations were run towards quasi-stationary states whose
energetic and helicity characteristics as mentioned above are roughly equal to the decaying 
run.
The MHD magnetic Prandtl number $\mathsf{Pr_m}=\mu/\eta$ is
unity. The Reynolds numbers of all configurations are of order $10^3$.

Three cases are considered. Setup (a) represents decaying
macroscopically isotropic 3D MHD turbulence.  The dataset contains $9$
states of fully developed turbulence each comprising $1024^3$ Fourier
modes. The samples are taken equidistantly in time over a period of about $3$ large
eddy turnover times.  The angle-integrated energy spectrum of this
system exhibits a Kolmogorov-like scaling law \cite{kolmogorov:k41a}
in the inertial range, i.e. $E_k\sim k^{-5/3}$.  The second dataset
(b) contains simulation data of a driven quasi-stationary
macroscopically anisotropic MHD flow with a strong constant mean
magnetic field. The driving is accomplished by freezing the largest
Fourier modes of the system ($k\le 2)$. The data comprises $1024 ^{2}$
Fourier modes perpendicular to the direction of the mean field and
$256$ modes parallel to it.  This dataset covers about $2$ large eddy
turnover times of quasi-stationary turbulence with $8$ samples taken
equidistantly over that period.  The perpendicular energy spectrum
shows Iroshnikov-Kraichnan-like behavior $E_{k_\perp}\sim
k_\perp^{-3/2}$ \cite{iroshnikov:ikmodel,kraichnan:ikmodel}. Note that this 
is neither claim nor clear evidence for the validity of the Iroshnikov-Kraichnan picture in 
this configuration. For
further details of the simulations and additional references see
\cite{mueller_grappin:endyn}.  The third simulation (c) represents a
turbulent statistically isotropic Navier-Stokes flow with resolution $1024^3$ which is kept
stationary by the same driving method as in case b) and exhibits Kolmogorov-scaling 
$E_k\sim k^{-5/3}$ of the turbulent energy spectrum.

For all turbulent systems the statistical
properties of $\delta f(\ell)$ which is computed  over varying scale $\ell$
are investigated.  In the present work $f$ stands for the component of
$\vc{z}^+$ in the increment direction $\mathbf{\hat{e}}$ or the fluctuation
energy defined here as $E\equiv (z^+)^2$. 
For the macroscopically isotropic setups (a) and (c) $\mathbf{\hat{e}}=\mathbf{\hat{e}}_z$. 
In case (b) the
unit vector points in a fixed arbitrary direction perpendicular to the
mean magnetic field.  Under the assumption of statistical isotropy
which is approximately fulfilled in setup (a), (c), and in system (b)
in planes perpendicular to the mean magnetic field, the statistical
properties of $\delta f(\ell)$ depend solely on $\ell$. This assumption also holds approximately 
for $\delta E$ if contributions by eddies on larger-scales which are convolved into this non-Galileian-invariant quantity can be regarded as quasi-constant on scale $\ell$ (see below). 
The quantity $\langle \delta f(\ell)\rangle$ scales
self-similarly with the scaling parameter $\alpha$ ($\alpha \geq 0$), if
$\langle \delta f(\lambda \ell)\rangle=\lambda^\alpha \langle f(\ell)\rangle$ 
for every $\lambda$.
For the associated cumulative probability distribution follows  
$\wp(\delta f(\ell)\leq \rho)=\wp(\lambda^{-\alpha} \delta
f(\lambda\ell) \leq \rho)$ for any real $\rho$. 
This implies for the probability density $P$ 
\begin{equation}
P[\delta f(\ell)] =\lambda^{-\alpha}P_s[\lambda^{-\alpha}\delta f_s]\label{possible}
\end{equation}
introducing  the master PDF $P_s$ with $\delta f_s=\delta f(\lambda \ell)$. 
According
to Eq. (\ref{possible}), there is a family of PDFs that can be collapsed
to a single curve $P_s$, if $\alpha$ is independent of $\ell$. This is known
as monoscaling in contrast to multifractal scaling observed, e.g., for two-point increments
of a turbulent velocity field.

\begin{figure}
\centerline{\includegraphics[width=0.5\textwidth]{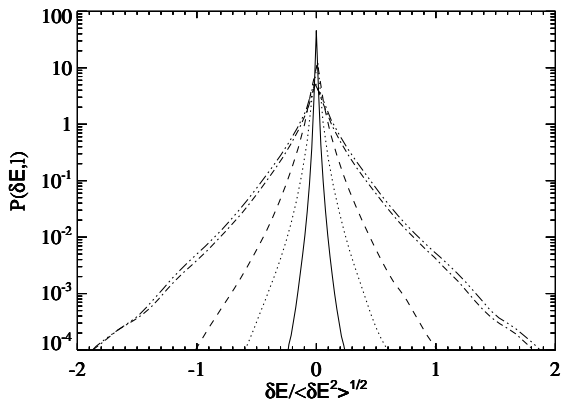}}
\caption{The PDFs of total energy fluctuations $ \delta E$ on five different scales 
$\ell=\pi/n$ with $n=511$ (solid), $n=130$ (dotted), $n=46$ (dashed), $n=4$ (dot-dashed), $n=1$ (3-dot-dashed).}
\label{f1} \end{figure}

To test if the abovementioned observations in the solar
wind are a phenomenon related to inherent properties of turbulence
time- and space-averaged increment series $\delta
z^{+}(\ell)$ and $\delta E(\ell)$  for
different $\ell$, ranging between $\pi/512$ up to $\pi$ are computed.
In system (a) the
increments are normalized using  $(E^\mathrm{T})^{1/2}$ with 
$E^\mathrm{T}=1/4\int_V\mathrm{d}V[(z^+)^2+(z^-)^2]$ 
to compensate for the decaying amplitude of the
turbulent fluctuations.  The PDFs are generated as normalized
histograms of the respective increments taken over all positions in
the $2\pi$-periodic box which contains the real space fields,
$\vc{v}(\vc{r})$ and $\vc{b}(\vc{r})$, computed from the available
Fourier-coefficients.  Fig. \ref{f1} shows
$P[\delta E(\ell)]$ for various $\ell$ in the isotropic
case (a).  The non-Gaussian nature of the PDFs over all scales is
evident.  Similar behavior is found in the anisotropic case (b) 
where the increments are taken perpendicularly to the direction
of the mean field as well as in the Navier-Stokes simulation (c).  
The PDFs are highly symmetric and become
increasingly broader with growing $\ell$ reflecting the increase of
turbulent energy towards largest scales.  Interestingly, the PDFs at \textit{all}
scales have the same leptokurtic shape resembling
L\'evy laws. In particular, away from the center, $\delta E=0$, 
the PDFs are close to gamma distributions $\sim \exp(-|\delta E|/\Delta)|\delta E|^{-\gamma}$ of 
different widths $\Delta$.
The exponent $\gamma$ of the best fits is constant in the inertial range and amounts 
approximately to $3.4$ (a), $4.2$ (b), and $3.1$ (c).  
In the solar wind a similar finding however with $\gamma\approx 2.5$ was reported \cite{hnat_etal:monoscalsolar}.
 
The similarity of the $P[\delta E(\ell)]$ on different scales $\ell$ suggests the possibility of monoscaling.
The monoscaling exponent
is expected to be scale-independent in the inertial range only since the energy increments are not Galilei invariant.
Therefore, small-scale $\delta E$ also comprise contributions by larger eddies which advect the small-scale fluctuations.
A linearization of $\delta E$ with respect to the largest-scale contribution $(z^+_0)^2\gg (\delta z^+)^2$ yields to lowest order 
$\delta E\approx 
(z_0^++ \delta z^+)^2\sim z^+_0\delta z^+$. As a consequence, the energy increments reflect the 
inertial-range scaling  of the turbulent Els\"asser fields, i.e. $\delta E\sim\delta z^+\sim \ell^\alpha$.
To apply the rescaling procedure given by Eq. (\ref{possible})
(cf. also \cite{hnat_etal:monoscalsolar}) the exponent $\alpha$ is
extracted from the PDFs by two independent techniques. 

\begin{figure}
\centerline{\includegraphics[width=0.5\textwidth]{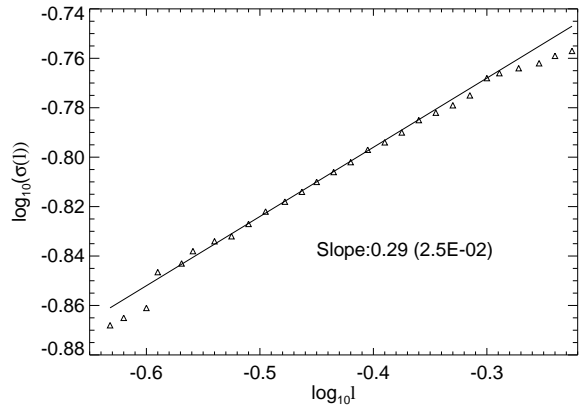}}
\caption{Standard deviation of total energy increments  within the inertial range in case (a) (triangles) with 
 linear least-squares fit (solid line).}
\label{f3} \end{figure}

Firstly, the standard deviation is considered which is defined as 
$\sigma(\ell)=[\langle\delta E(\ell)^2\rangle]^{1/2}$. In the inertial range $\sigma$ exhibits power-law behavior
with respect to the increment distance, $\sigma(\ell)\sim\ell^\alpha$,
Fig. \ref{f3} 
shows the standard deviation of total energy fluctuations in the inertial range
for the isotropic case (a) in double logarithmic
presentation. A linear least-squares fit 
is carried out to obtain $\alpha$. The
characteristic exponents deduced in this way are $ \alpha=0.29 \pm
0.025$ for the isotropic case (a), $\alpha=0.23 \pm 0.025$ for the
anisotropic case (b), and $\alpha=0.28 \pm 0.03$ for the
Navier-Stokes flow (c).  As expected these values are close to
the non-intermittent scaling exponents observed for the turbulent
field fluctuations, i.e.  $\alpha_\mathrm{K41}=1/3$ for cases (a) and (c) while
$\alpha_\mathrm{IK}=1/4$ for case (b).

\begin{figure}
\centerline{\includegraphics[width=0.5\textwidth]{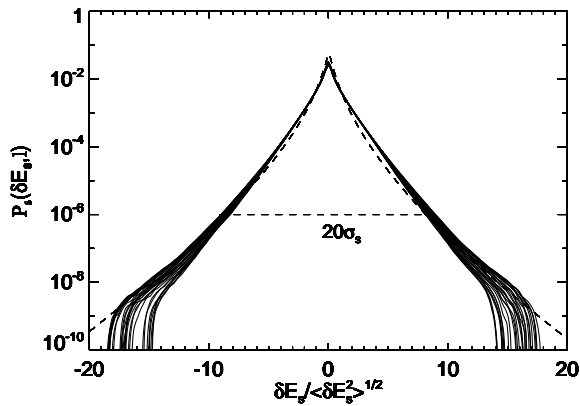}}
\caption{Rescaled PDFs of total energy fluctuations in the inertial range of
the Navier-Stokes case (c). The gamma law $10^{-3}\exp(-|\delta E|/0.4)|\delta E|^{-3.4}$ is represented by the dashed curve.}
\label{f5} \end{figure}

Secondly, in the inertial range the characteristic exponents can be obtained via the
amplitude of $P(0,\ell)\sim \ell^{-\alpha}$ 
profiting from the fact that the peaks of the PDFs are statistically
the least noisy part of the distributions. The scaling
exponent obtained by using this method is in good agreement with the
value of $\alpha$ obtained via the PDF variance.
\begin{figure}
\centerline{\includegraphics[width=0.5\textwidth]{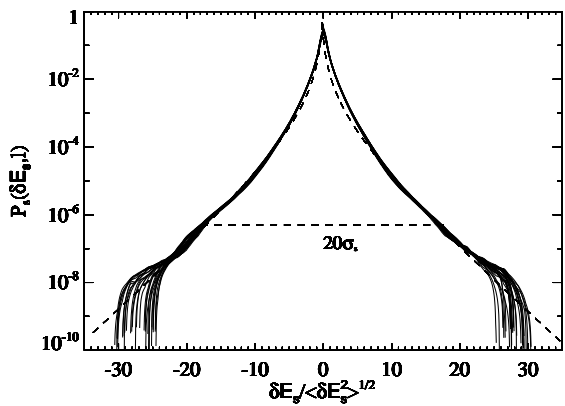}}
\caption{Rescaled PDFs of total energy fluctuations in the
inertial range of the isotropic case (a). 
The gamma law $10^{-3}\exp(-|\delta E|/0.35)|\delta E|^{-3.1}$ is represented by the dashed curve.}
\label{f4} \end{figure}
Fig. \ref{f4} shows the rescaled PDFs according to Eq. (\ref{possible}) for MHD case (a) (similar for (b), not shown)
while Fig. \ref{f5} displays the rescaled PDFs obtained from the Navier-Stokes simulation (c). The corresponding increment
distances $\ell$ are all lying in the 
respective inertial range. Evidently the PDFs are self-similar and collapse for up to 
$20\sigma$ with weak scattering on the master PDF, $P_s$, when using the characteristic exponents given above. The dashed lines in both figure display the best fitting gamma laws.
%
 
\begin{figure}
\centerline{\includegraphics[width=0.5\textwidth]{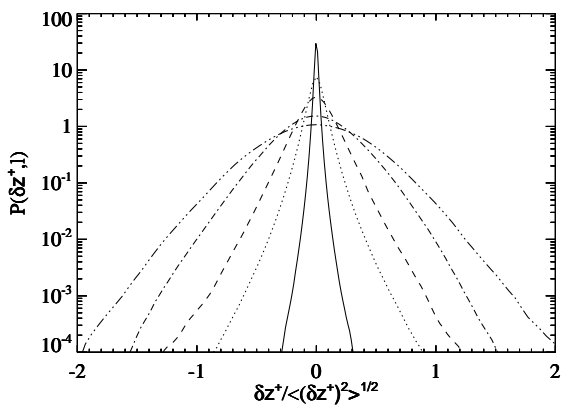}}
\caption{The PDFs of Els\"asser field fluctuations $ \delta z^{+}$ for the same five different scales as in Fig. 
\ref{f1}.}
\label{f2} \end{figure}
The PDFs of the Els\"asser field fluctuations, P[$\delta z^+(\ell)]$, in system (a)
(systems (b) and (c) likewise) display a different and
well-known behavior as can be seen from Fig. \ref{f2}.
The distributions lose their small-scale leptocurtic character as
$\ell$ increases. Due to the lacking correlation of distant turbulent fluctuations the associated distributions 
become approximately Gaussian at large scales.
Because of the resulting multifractal scaling of
the PDFs which is a signature of the intermittent small-scale
structure of turbulence 
it is obvious that they can not
be collapsed onto a single curve even in the inertial range. 
However,
one can infer the nonintermittent characteristic scaling exponent by
regarding the function $P(0,\ell)$ (not shown).  For example, in system (a) this function
exhibits clear inertial-range scaling $\sim \ell^{-a}$ with $a=
0.33\pm 1.5\times10^{-2}$ in very good agreement with $\alpha_\mathrm{K41}$. 

The occurrence of gamma PDFs is made plausible by a simple reaction-rate ansatz 
\cite{cheng_redner:fragmodel,sornette:gammaexpl}:
Consider the `intensity' $n(e)$ of turbulent fluctuations with energy
$e=|\delta E|$ such that $n(e)$ is the  
fraction the total turbulent energy associated with these fluctuations and the larger
eddies in which they are embedded.
The evolution of this function is assumed to obey the 
following linear rate equation:
\begin{equation}
\partial_t n(e) = - n(e)/\tau_-(e) +\int_e^\infty \mathrm{d}e' n(e')
/\tau_+(e',e) \label{fragmodel}
\end{equation}
where $\tau_+(e',e)$ is the time characteristic of the creation of fluctuations with energy $e$  
as a result of
turbulent transfer from fluctuations with energy $e'$ while $\tau_-(e)$ is the respective characteristic decay time.
Normalization of 
$ n(e)$ by $\int_0^\infty\mathrm{d}e' n(e')$
yields the corresponding PDF $P(e)$. 
In a statistically 
stationary state Eq. (\ref{fragmodel}) 
then gives
\begin{equation}
P(e)=C_1\int_e^\infty\mathrm{d}e' P(e')\frac{\tau_-(e)}{\tau_+(e',e)}\label{gammmodel}
\end{equation}
where $C_1$ is a normalization constant. For $\tau_-(e)/\tau_+(e',e)\sim
 (e'/e)^{\gamma}$ this integral equation has the
solution $P(e)=C_2 e^{-\gamma}\exp(-e/\Delta)$.
Thus, the model (\ref{fragmodel}) which mimics in combination with the abovementioned assumptions
a direct spectral
transfer process yields the 
observed gamma distributions.
Note that the lower bound of the integral in 
Eq. (\ref{gammmodel}) implies that 
energy flows from higher to lower levels where for technical simplicity very large differences between $e$ and $e'$ are allowed.
A finite upper bound of the integral
in Eq. (\ref{gammmodel}) does, however,  not change the result fundamentally.
This suggests that the observed gamma distributions are an indication of  
turbulent spectral transfer. 

In summary it has been shown by high-resolution direct numerical simulations
of incompressible turbulent magnetohydrodynamic and Navier-Stokes flows
that the monoscaling of energy fluctutation PDFs 
observed in the solar wind 
is the consequence of lacking Galilei invariance of energy increments in combination with 
self-similar scaling of the underlying turbulent fields. 
The closeness of the PDFs to L\'evy-type gamma distributions is made plausible 
by a simple model mimicking nonlinear spectral transfer.


\acknowledgments
M.M. thanks the Max-Planck-Institut f\"ur Plasmaphysik where this work
was carried out for its hospitality. The authors thank A. Busse for
supplying the raw numerical Navier-Stokes data.

\newcommand{\nop}[1]{}

\end{document}